\documentclass[12pt,reqno]{amsart}

\usepackage{amsmath}
\usepackage{amsfonts}
\usepackage{amssymb}
\usepackage{latexsym}   
\usepackage{graphicx}
\usepackage{color}
\usepackage{hyperref,subfig}
\usepackage{algorithm,algorithmic}

\usepackage{vmargin,fancyhdr}   

\newcommand{\beql}[1]{\begin{equation}\label{#1}}
\newcommand{\eeq}{\end{equation}}

\newtheorem{theorem}{Theorem}

\setpapersize{USletter}
\setmarginsrb{.75in}{.5in}        
             {.75in}{.5in}        
             {.25in}{.25in}     
             {.25in}{.5in}      
\newcommand{\hide}[1]{}

\newcommand{\beq}[1]{\begin{equation}\label{#1}}

\newcounter{rot}

\newcommand{\ignore}[1]{}

\title{Perfect Reconstruction of Oncogenetic Trees}

\author{Charalampos E. Tsourakakis}
\address{Department of Mathematical Sciences\\
Carnegie Mellon University\\
5000 Forbes Av., 15213\\
Pittsburgh, PA \\
U.S.A} \email{ctsourak@math.cmu.edu}

\begin{document}

\maketitle

\begin{abstract}
In this note we provide the necessary and sufficient conditions to uniquely reconstruct an oncogenetic tree. 
\end{abstract}

\section{Introduction} 
\label{sec:intro}

Human cancer is caused by the accumulation of genetic alternations in cells \cite{michor,weinberg}.
Finding driver genetic mutations, i.e., mutations which confer growth advantage on the cells carrying them and have been positively selected during the evolution of the cancer
and uncovering their temporal sequence have been central goals of cancer research the last decades \cite{nature}.
Among the triumphs of cancer research stands the breakthrough work of Vogelstein and his collaborators \cite{fearon,vogelstein} 
which provides significant insight  into the evolution of colorectal cancer. 
Specifically, the so-called ``Vogelgram'' models colorectal tumorigenesis as a linear accumulation of certain genetic events. 
Few years later, Desper et al. \cite{desper} considered more general evolutionary models compared to 
the ``Vogelgram'' and presented one of the first theoretical approaches to the problem \cite{michor},
the so-called {\it oncogenetic trees}. 
Before we provide a description of oncogenetic trees which are the focus of our work,
we would like to emphasize that since then a lot of research work has followed from several groups of researchers,
influenced by the seminal work of Desper et al. \cite{desper}. 
Currently there exists a wealth of methods that infer evolutionary models from 
microarray-based data such as gene expression and array Comparative Genome Hybridization (aCGH) data: 
distance based oncogenetic trees \cite{desper2}, maximum likelihood oncogenetic trees \cite{heydebreck},
hidden variable oncogenetic trees \cite{tofigh}, conjuctive Bayesian networks \cite{beer4}
and their extensions \cite{beer6,beer3}, mixture of trees \cite{beer5}.
The interested reader is urged to read the surveys of Attolini et al. \cite{michor} and Hainke et al. \cite{hainke}
and the references therein on established progression modeling methods. 
Furthermore, oncogenetic trees have successfully shed light into many types of cancer such as renal cancer \cite{desper}, 
hepatic cancer \cite{hepatic} and head and neck squamous cell carcinomas \cite{huang}.

\subsubsection*{Oncogenetic Trees} 
An oncogenetic tree is a rooted directed tree\footnote{Typically, the term 
{\it tree} is reserved for the undirected case and the term {\it branching} for the directed
case. Throughtout this note, we consistently 
use the term {\it tree} mean a directed tree as in \cite{desper}.}. The root  represents the state of tissue with no mutations. 
Any other vertex $v \in V$ represents a mutation. Each edge represents a ``cause-and-effect'' relationships.  
Specifically, for a mutation represented by vertex $v$ to occur, all the mutations corresponding to vertices that lie on the directed path 
from the root to $v$ must be present in the tumor.
In other words, if two mutations $u,v$ are connected by an edge $(u,v)$ then $v$ cannot occur if $u$ has not occured. 
The edges are labeled with probabilities.
Each tumor corresponds to a rooted subtree of the oncogenetic tree and the probability of occurence is determined 
as described by \cite{desper}. Desper et al. provide an algorithm that finds a likely oncogenetic tree that fits 
the observed data. 

In this work we answer a fundamental question regarding oncogenetic trees. Before we state the question, we introduce some notation first. 
Let $T=(V,E,r)$ be a rooted tree on $V$, i.e., every vertex has in-degree at most one 
and there are no cycles, and let $r \in V$ be the root of $T$.
Given a finite family $\mathcal{F}=\{A_1,...A_q\}$ of sets of vertices, i.e., $A_i \subseteq V(T)$ for $i=1,\ldots,q$,
where each $A_i$ is the vertex set of a $r$-rooted sub-tree of $T$, what are the necessary and sufficient conditions, if any,  
which allow us to uniquely reconstruct  $T$? 
In this work we treat this natural combinatorial question, namely:

\begin{center}
{\em ``When can we reconstruct an oncogenetic tree $T$ from a set family $\mathcal{F}$?''}
\end{center}

Despite the fact that in practice aCGH data tend to be noisy and consistent with more than one oncogenetic trees, 
the question is nonetheless interesting and to the best of our knowledge 
remains open so far \cite{papadimitriou,schaeffer}.
Theorem~\ref{thrm:thrm1} provides the necessary and sufficient conditions to uniquely reconstruct 
an oncogenetic tree. 
We write $u \prec  v$ ($u \nprec v$) to denote that $u$ is (not) a descendant of $v$ in $T$.

\begin{theorem}
\label{thrm:thrm1}
Let $T$ be an oncogenetic tree and $\mathcal{F}=\{A_1,...A_q\}$ be a finite family of sets of vertices, 
i.e., $A_i \subseteq V(T)$ for $i=1,\ldots,q$, where each $A_i$ is the vertex set of a $r$-rooted  sub-tree of $T$ 
The necessary and sufficient conditions to uniquely reconstruct the tree $T$ from the family $\mathcal{F}$ are
the following:
\begin{enumerate}
		\item For any two distinct vertices $x,y \in V(T)$ such that $(x,y) \in E(T)$, 
                there exists a set $A_i \in \mathcal{F}$ such that $x \in A_i$ and $y \notin A_i$.
	        \item For any two distinct vertices $x,y \in V(T)$ such that $y \nprec x$ and $x \nprec y$
	        there exist sets $A_i,A_j \in \mathcal{F}$ such that $x \in A_i$, $y \notin A_i$ and $x \notin A_j$ and $y \in A_j$.
\end{enumerate} 
\end{theorem}

\noindent In Section~\ref{sec:proof} we prove Theorem~\ref{thrm:thrm1}. It is worth noticing that our proof 
provides a simple procedure for the reconstruction as well. 

\section{Proofs} 
\label{sec:proof}

\noindent 
In the following we call a tree $T$ {\it consistent} with the family set $\mathcal{F}$ if  
all sets $A_i \in \mathcal{F}$ are vertices of rooted sub-trees of $T$. 
Notice that when two or more trees are consistent with the input dataset $\mathcal{F}$, 
then we cannot uniquely reconstruct $T$. 

\begin{proof} 
First we prove the necessity of conditions 1,2 and then their sufficiency to reconstruct $T$. 

\begin{table}
\centering
\label{tab:tab1} 
  \begin{tabular}{|c|c|} \hline  
   \includegraphics[width=0.4\textwidth]{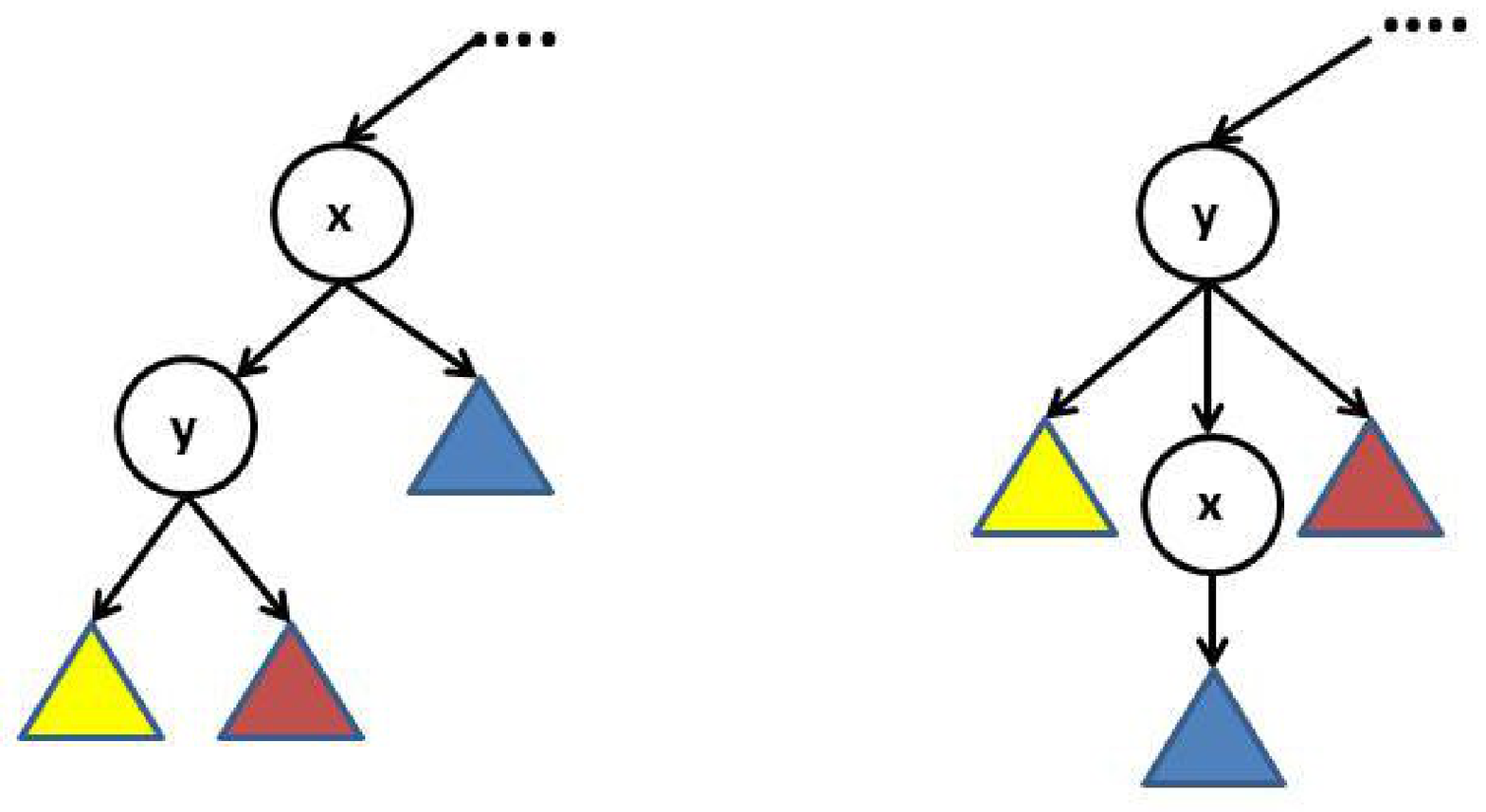} & \includegraphics[width=0.4\textwidth]{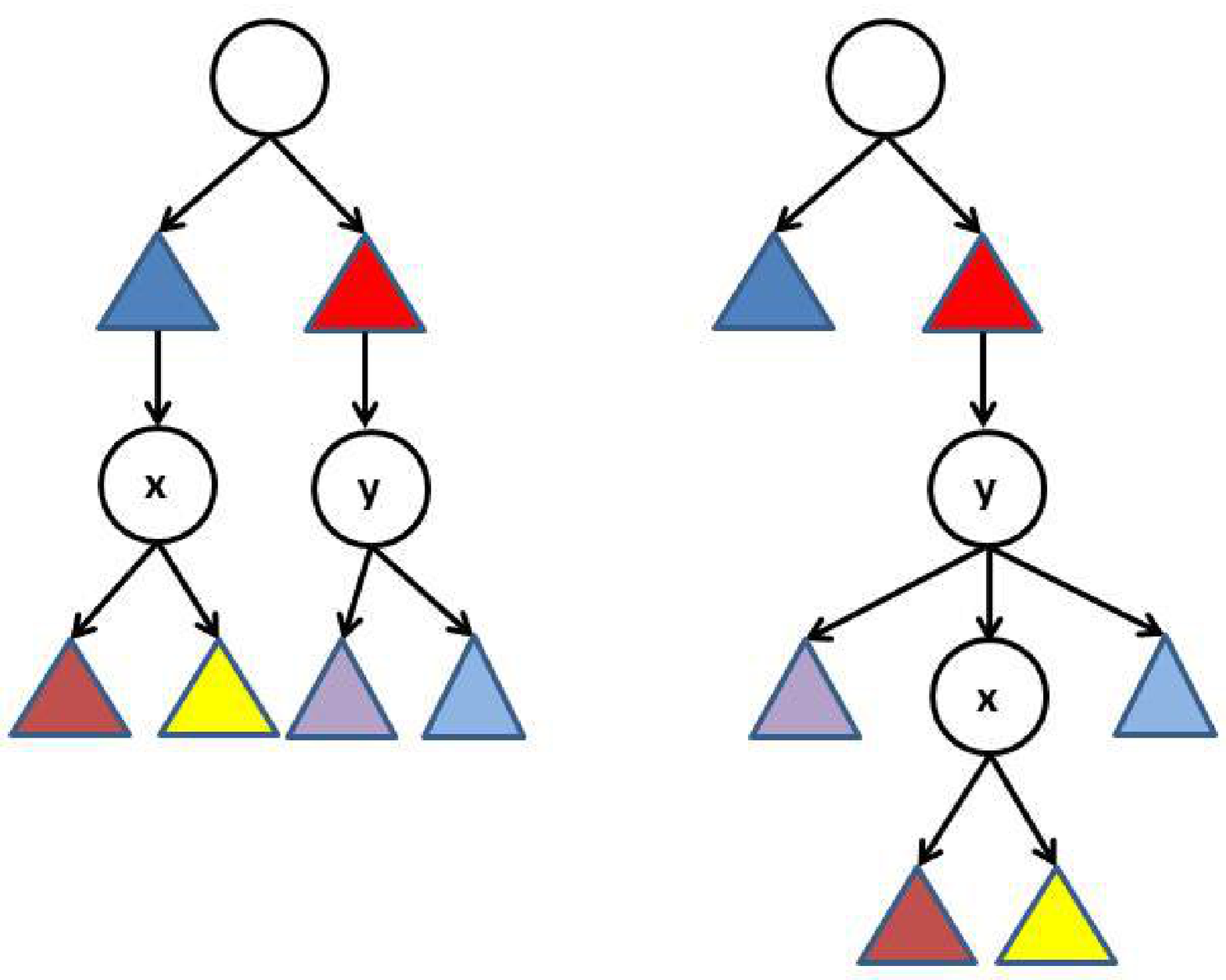}  \\ \hline
   Figure 2(a) Necessity of Condition 1 & Figure 2(b) Necessity of Condition 2  \\ \hline
  \end{tabular}
  \caption{Illustration of necessity conditions of Theorem~\ref{thrm:thrm1}.}
\end{table}

\underline{Necessity:} For the sake of contradiction, assume that Condition 1 does not hold. Therefore, 
there exists two vertices $x,y \in V(T)$ such that there exists no set $A \in \mathcal{F}$ 
that contains one of them.  Then, the two trees 
shown in Figure 2(a) are both consistent with $\mathcal{F}$. Therefore we cannot reconstruct $T$. 
Similarly, assume that Condition 2 does not hold. Specifically assume that 
for all $j$ such that $x \in A_j$, then $y \in A_j$ too (for the symmetric case
the same argument holds). Then, both trees in Figure 2(b) are consistent with $\mathcal{F}$ 
and therefore $T$ is not reconstructable. The symmetric case follows by the same argument. 

\underline{Sufficiency:} Let $x \in V(T)$  and $P_x$ be the vertex set of the unique path from the root $r$ to $x$, i.e., $P_x=\{r,\ldots,x\}$. 
Also, define $F_x$ to be the intersection of all sets in the family $\mathcal{F}$ that contain vertex $x$,
i.e., $F_x =\underset{\mbox{ $A_i \ni x$ }}{ \text{~~}\bigcap A_i }$. 
We prove that $F_x = P_x$. Since by the definition of an oncogenetic tree $P_x \subseteq F_x$ it suffices
to show that $F_x \subseteq P_x$. Assume for the sake of contradiction that $F_x \nsubseteq P_x$.
Then, there exists a vertex $v \in V(T)$ such that $v \in F_x, v \notin P_x$. We consider the following three cases. 

\noindent 
\underline{$\bullet$ {\sc Case 1} ($x \prec v$):} Since by definition each set $A_i \in \mathcal{F}$ is the vertex 
set of a rooted sub-tree of $T$,  $v \in P_x$ by the definition of an oncogenetic tree.

\noindent 
\underline{$\bullet$ {\sc Case 2} ($v \prec x$):} Inductively by condition 1,
there exists $A_i \in \mathcal{F}$ such that $x \in A_i, v \notin A_i$. Therefore, $v \notin F_x$. 

\noindent 
\underline{$\bullet$ {\sc Case 3} ($x \nprec v, v \nprec x$):}
By condition 2, there exists $A_i \in \mathcal{F}$ such that  $x \in A_i$ and $v \notin A_i$.
Hence, $v \notin F_x$. 

In all three cases above, we obtain a contradiction and therefore $v \in F_x \Rightarrow v \in P_x$.
Therefore, $F_x \subseteq P_x$ and subsequently $F_x = P_x$. 
Given this fact, it is easy to reconstruct the tree $T$. We sketch the algorithm: compute for each $x$ the set 
$F_x$ which is the unordered set of vertices of the unique path from $r$ to $x$.
The ordering of the vertices which results in finding the path $P_x$, i.e.,
$(v_0=r \rightarrow v_1 \rightarrow .. v_{k-1} \rightarrow v_k=x)$ is computed using sets in $\mathcal{F}$ 
which contain $v_i$ but not $v_{i+1}$, $i=0,..,k-1$. The existence of such sets is guaranteed by condition 1.  
\end{proof}

\section{Acknowledgments} 

Research supported by NSF Grant No. CCF-1013110. The author thanks Deepak Bal and Professor Alan Frieze for their feedback.

\end{document}